\documentclass[12pt,preprint]{aastex}

\def\kms{\relax \ifmmode {\ \rm km s}^{-1}\else \ km\ s$^{-1}$\fi}

\def\Mso{{M$_{\rm \odot}$}}

\def\cm3{${\rm cm}^{-3}~$}

\def\nii{[N~{\sc ii}]}

\def\heii{He~{\sc ii}}
\def\oiii{[O~{\sc iii}]}

\def\ha{H$\alpha~$} 
\def\hb{H$\beta$}
 
\slugcomment{}

\shorttitle{CSs of PNe in the Small Magellanic Cloud}
\shortauthors{Villaver, et al.}
  
\begin{document}
  
\title{The Low- and Intermediate-Mass Stellar
Population in the Small Magellanic Cloud: The Central Stars of Planetary
Nebulae\footnote{Based on observations made with the NASA/ESA Hubble Space
Telescope, obtained at the Space Telescope Science Institute, which is
operated by the Association of Universities for Research in Astronomy, Inc.,
under NASA contract NAS 5--26555}}

\author{Eva Villaver\altaffilmark{2}}
\affil{Space Telescope Science Institute, 3700 San Martin Drive,
Baltimore, MD 21218, USA; villaver@stsci.edu}
\author{Letizia Stanghellini\altaffilmark{3}, and Richard A. Shaw}
\affil{National Optical Astronomy Observatory, 950 N. Cherry Av.,
Tucson, AZ  85719, USA; letizia@noao.edu, shaw@noao.edu}

\altaffiltext{2}{Affiliated with the Hubble Space Telescope Division
of the European Space Agency} 
\altaffiltext{3}{On leave from INAF-Bologna Observatory}
\begin{abstract}

We present a study on the central stars (CSs) of Planetary Nebulae (PNe)
observed in the Small Magellanic Cloud (SMC) with the {\it Space Telescope
Imaging Spectrograph} instrument on-board the HST. The stellar magnitudes
have been measured using broad-band photometry, and Zanstra analysis of the
nebulae provided the stellar temperatures. From the location of the CSs on
the HR diagram, and by comparing the observed CSs with current models of
stellar evolution, we infer the CSs masses. We examine closely the
possibility of light contamination in the bandpass from an unrecognized
stellar companion, and  we establish strong constraints on the existence and
nature  of any binary companion. We find an average mass of 0.63~\Mso, which
is similar to the mass obtained  for a sample of CSs in the LMC
(0.65~\Mso). However, the SMC and LMC CS mass distributions differ slightly,
the SMC sample   lacking an intermediate-mass stellar population (0.65 to
0.75~\Mso).  We discuss the significance and possible reasons for the
difference  between the two mass distributions.  In particular, we consider
the differences in the star formation history  between the clouds and the
mass-loss rate dependence on metallicity.

\end{abstract}

 
\section{INTRODUCTION}
Central Stars (CSs) of Planetary Nebulae (PNe) are the result of the
evolution of low- and intermediate-mass stars (1 to $\sim$5~\Mso) that
loose most of their stellar envelope through mass-loss towards the end of the
Asymptotic Giant Branch (AGB) phase. 

The CS mass depends mainly on the stellar mass during the Main-Sequence (MS)
phase (hereafter, the initial mass), and on the mass-loss during the
AGB. Neither the process responsible for the 
high mass-loss rates during the AGB phase nor its dependency on metallicity
are well understood. In the commonly accepted scenario the mass-loss is
thought to be driven mainly by a combination of two processes: dust formation
by shock waves
caused by the Mira-like stellar pulsation, and the acceleration of dust by
radiation pressure \citep{Wod:79,Bow:88}. The transfer of momentum from the
dust to the gas ultimately drives the outflow.  

The dust formation process depends on
the chemical composition of the gas: the lower the metallicity, the smaller
the amount of dust formed,  and the lower the efficiency of the momentum
transfer to the gas. Thus, low metallicity stars with dust-driven winds are
expected to loose smaller amounts of matter \citep{Wetal:00}. \cite{Wil:00}
has shown that mass-loss during the AGB phase can occur even in the absence
of dust, but its efficiency is then much lower. 
In the case where the mass-loss occurs in the absence of dust, 
low metallicity stars are expected  to have 
reduced mass-loss rates because  they have a
smaller radius for a given mass and luminosity
\citep{Wbs:96,Wil:00}. Thus, everything, seems to conspire
against mass-loss at low metallicity.  
As a consequence, a higher fraction of main sequence stars should
reach the Chandrasekhar mass limit in the SMC than in the 
LMC \citep{Uetal:99,Detal:99,Getal:00}, implying a higher rate of type II
Supernovae explosions in the SMC than in the LMC, if we assume identical
initial, low- and intermediate-mass stellar populations in the two Clouds.

Studies of CSs in the Galaxy are hampered by the poor knowledge of their
distances, a problem that can be overcome by observing PNe in the Magellanic
Clouds (LMC and SMC) \citep{Sta:00}. The high spatial resolution 
capabilities of the
Hubble Space Telescope ({\it HST}) is perfectly matched to resolve the
Magellanic Cloud PNe, and to directly observe their CSs. Furthermore,  LMC
and SMC CSs are affected by low field reddening. Since the metallicity of
the LMC and SMC are on average half and quarter that of the solar mix
\citep{Rb:89,Rd:90}, studying samples of CSs in both Clouds allows one 
to directly probe stellar evolution and mass-loss with different
metallicities. 
In this paper, we  determine the mass of a sample of CSs in the SMC. The aim
is to compare  them with a similarly-selected sample  of LMC CSs
(\citealt{Vss:03}; hereafter Paper I) to explore the effect of a different
metallicity on the CS mass distribution. The goal of this series of
papers is to determine accurate CSs masses of a large number of
extragalactic PNe, and to ultimately relate the
final mass to the initial mass of these sub-Chandrasekhar mass stars.

We present {\it Space Telescope Imaging Spectrograph} (STIS) broad-band
imaging of 27 targets in the SMC. In \S2 we describe the
observations. In \S3 we present the methodology and the results of the stellar
photometry, the Zanstra temperature derivation, the
bolometric correction, and the derivation of the stellar luminosities; in \S4
we discuss the correlations and the statistics of the measured stellar
parameters; finally, in \S5 we discuss and summarize our findings.

\section{OBSERVATIONS}

The observations of the 27 targets presented in this paper are from the {\it
HST} SNAPSHOT program 8663. The observations were made in direct imaging
(50CCD) and in slitless  mode with the G430M and G750M gratings of
STIS. The observing log, observing
configuration, target selection, acquisition, and basic calibrations 
are described in \cite{Setal:03}, as well as the morphological
classification of the nebulae, the line intensities, and other nebular
diagnostics obtained from the slitless spectra. In this paper we present the
photometric analysis performed on the STIS broad-band images
(50CCD). The 50CCD has sensitivity from $\sim$ 2,000 to 10,300 \AA~and a
spatial scale of 0\arcsec.0507 pix$^{-1}$. The FWHM is close to 2 pixels at
the central wavelength of the 50CCD   5850 \AA~\citep{Letal:01}. The
observations were made with the CCD detector using a gain of 1 e$^-$  per
analog-to-digital converter unit. All the exposures were split into two equal
components to facilitate cosmic-ray removal.  In Table~1 we give in column
(1) the PN name (we favor the SMP nomenclature, when available);  column
(2) the total integration time, and in column (3) we note whether the
CS has been detected in the image.

\section{ANALYSIS}
\subsection{Stellar Photometry}

The technique to measure the CS photometry has already been described in
detail in Paper I. To summarize it, we have applied aperture photometry
using the IRAF\footnote{IRAF is distributed by the National Optical Astronomy
Observatory, which is operated by the Association of Universities for
Research in Astronomy, Inc., under cooperative agreement with the National
Science Foundation.} {\bf phot} task. We have considered a stellar aperture
with a radius of 2 pixels to measure the flux of the CS. The contribution
from the nebula in the stellar aperture has been subtracted by estimating the
median nebular flux in an annulus with 
a width of 2 pixels adjacent to the stellar 
aperture. At the distance of the SMC, the {\it HST} spatial resolution
allowed us to resolve the nebula and separate the nebular and stellar
emission. The variation of the nebular flux within the annulus, 
is reflected in the standard deviation and is 
propagated in the errors of the measured magnitudes. The fraction of the
stellar point spread function (PSF) that falls  out of the stellar
aperture has been corrected using an aperture correction determined from
the curve of encircled energy derived by \cite{Betal:02} for stars near the
field center.

There are four unresolved objects (i.e. objects that have nebular extension 
similar to the stellar PSF) in our sample, J~23, SMP~1, SMP~18, and SMP~20. 
In these cases, the nebular
morphology is not available, however, we can still detect
the CS and measure its magnitude in all of the PNe except
SMP~18. Among the
other PNe, J~18, SMP~6, and SMP~24 are the most compact, yet spatially
resolved. As a consequence, the nebular contribution to the continuum that we
subtract  
to obtain the stellar magnitude is high and it has a small standard deviation
which is reflected in the reduced uncertainties in the measured
magnitude. Thus it should be noted that the  
magnitude errors given for J~18, SMP~6 and SMP~24 are probably underestimated.

When the CS is detected at a very low level above the nebular emission, or
the nebular emission decreases very sharply with the distance from the star
we can considerably reduce the systematic photometric errors by subtracting a
nebular image before performing the photometry. The method has been described
in detail in Paper I. The nebular image is built by stacking the
monochromatic images,  \hb, \oiii~4959, 5007~\AA, \ha, and \nii~6548 and
6584~\AA~lines, obtained from the slitless STIS spectra. In the case where
the CS is not detected in the 50CCD images, we have estimated a lower limit
to the CS magnitude by performing aperture photometry at the geometrical
center of the nebulae. 

The zero-point calibration to magnitudes in the STMAG\footnote{The STMAG is
the Space Telescope magnitude system, based on a spectrum  with constant flux
per unit wavelength.} system has been taken from \cite{Betal:02}. As in Paper
I, we have neglected the STIS charge transfer efficiency (CTE), since it has
been shown to be below 0.01 dex \citep{Retal:00} for stars near the field
center. The image distortions can also be neglected, since they are 
known to be negligible in this observing mode.

The magnitudes have been corrected for extinction using the amount of
nebular extinction derived from the Balmer decrement, and the  relation {\rm
c}~=~1.41~E$_{B-V}$ \citep{Sea:79}. The logarithmic  extinction constant at
\hb, {\rm c}, has been taken from \cite{Setal:03} except for J~18 
\citep{JK93}, J~23, and J~27 \citep{BL}, MG~13 \citep{Ld:03}, and SMP~20
\citep{Ld:96} for which we have used the values given in the literature
since no G430M exposures were available. 
We have used the interstellar extinction law of \cite{Sm:79}, 
and assumed that R$_{V}$~=~3.1 in order to derive the extinction, because
in our wavelength range, the SMC extinction law is very
similar to the Galactic extinction law \citep{How:83}.

We have modeled the transformation from instrumental magnitudes in the
STMAG system to standard V-band magnitudes in the photometric Johnson-Cousins
UBVI system with IRAF/STSDAS\footnote{STSDAS is a product of the Space
  Telescope Science Institute, which 
is operated by AURA for NASA.} task {\bf synphot}. The median of the V-50CCD
colors 
obtained for blackbodies between 30,000 and 300,000 {\rm K} with E$_{B-V}$
values appropriate for each source has been used to transform the STMAGs into
V-band magnitudes (see Paper I for details). The uncertainty in the
transformation has been added in quadrature to the error in the magnitude.

In Table~2, we have summarized the results of the photometry. Column (1)
gives the PN  name; columns (2) and (3) give the STMAG and standard V
magnitudes respectively, or their lower limits, as well as the associated
errors. The errors quoted in this Table include several sources:
the random error (e.g., the photon and read-out noise), the
systematic errors (from the CS fluxes, the sky subtraction, etc), and the
errors in the calibration. The observations of  J~4 and SMP~18 were
saturated, and are noted in the Table.  
The magnitudes derived from saturated data
have not been included in the analysis, the mass calculation and
the plots. The extinction constants used to determine the color excesses
needed to correct for extinction are listed in column (4).

\subsection{Effective Temperatures, Bolometric Corrections, and Stellar
  Luminosities} 
 
Of the different methods available in the literature to determine the stellar
temperature (e.g. Stoy's method; \citealt{Sto:33}) we have adopted the
Zanstra method \citep{Zan:31,Hs:66,Kal:83}. The Zanstra method relates the
total ionizing flux of the star (hereafter assumed to have a blackbody
spectrum) to the flux of a  recombination line (of hydrogen or helium) and to
the stellar continuum flux in the V band. The implicit assumption behind the
Zanstra method is that all the photons above the  Lyman limit of H or He$^+$
are absorbed within the nebula, and that each recombination results in a
Balmer photon. When the \heii~4686 \AA~line flux is available, we estimate
two values for the stellar temperature, which are 
based on the hydrogen and the helium recombination lines.

The data needed for the temperature calculation (\hb~fluxes, nebular radii,
and extinction constants) have been taken from \cite{Setal:03} when
available.   The \hb~fluxes and extinction constants for J~18, J23, and J~27
have been taken from the literature \citep{JK93, BL}. In the case of MG~13
the extinction constant has been taken from \cite{Ld:03} and the \hb~flux has
been computed from the \ha~flux given in \cite{Setal:03} by assuming case B
recombination from \cite{Ost:89}, and a nebular gas temperature of
10,000 {\rm K}. The \heii~4686~\AA~line fluxes have been taken from
\cite{BL}, \cite{Ld:96,Ld:03}, \cite{Dm:91a,Dm:91b}, \cite{VDM}, \cite{MBC},
and \cite{JK93}. In order to assure the best  results we have been very
conservative with the errors in the fluxes quoted by the references. In
Table~2 column (4) we have listed the 4686~\heii~flux line intensities
relative to \hb~=~100, not  corrected for extinction (and their errors),  and
in column (5) we list the reference code (see the legend in the note to the
Table) for the \heii~fluxes.

In order to compute the bolometric correction (BC), we have used the relation
between BC and T$_{eff}$ given by \cite{Vgs:96}. When available, the
\heii~Zanstra temperature has been used to determine the BC (unless it has
been derived from upper limits to the 4686~\heii~ flux), 
since they are the most reliable, otherwise the H~{\sc i} Zanstra T$_{eff}$
has been used. The CSs luminosities have been computed by adopting a distance to
the SMC of 58.29 {\rm Kpc} \citep{Wes:97}, and an absolute solar bolometric
magnitude of M$_{bol,\odot}$= 4.75 mag \citep{All:76}.

The SMC has an intrinsic line of sight depth between 4 and 16 {\rm kpc}
\citep{Hh:89,Hhg:89,Gh:91,Gh:92}, depending on the portion of the SMC
considered. In a more recent paper, \cite{Cetal:01} obtained a line of sight
depth between 6 and 12 {\rm kpc} by using 12 SMC clusters. The positions of
the CSs detected in our sample span a range of 400 {\rm pc} in right
ascension and 2 {\rm kpc} in declination with respect to the optical center
of the SMC. 
From the dispersion in the distance to the SMC derived by \cite{Cetal:01}
from the analysis of SMC clusters positions 
we have estimated an average line of sight depth of 5 {\rm kpc} for the 
CSs in our sample. The distance uncertainty introduced by this
depth in the SMC will systematically increase the $\log$ L errors by 0.02. We
therefore conclude that the error in the luminosity due to 
the 3-dimensional structure of the SMC is negligible. 

In Table~3 we give in column (1) the PN name; in columns (2) and (3) we give
the effective temperature (in units of 10$^3$ {\rm K}) derived from the
Zanstra method for the helium and hydrogen recombination lines
respectively. Column (4) gives the absolute visual  magnitude; column (5)
gives the stellar luminosity. The BC and its error, computed by
propagating the error in the temperature is given in column (6). The
morphological classification in column (7) is from \cite{Setal:03}. Column
(8) contains comments relevant to the magnitude measurement.

\section{RESULTS}

\subsection{Effective Temperatures}
In Figure~1 we compare the ratio of the stellar effective temperature derived
from the \heii~4686~\AA~[T$_Z$(\heii)] and the \hb~[T$_Z$(H)] recombination
lines (the Zanstra ratio) versus the \heii~Zanstra temperature. 
It should be noted that, although we
have computed a T$_Z$(\heii) upper limit using the upper limit \heii~4686~\AA~
flux for some of the objects (see Table 3), we do not include them in
the plot, or in the calculations. The Zanstra ratio approaches unity as
T$_Z$(\heii) increases, which is explained as an  
effect of the optical thickness of the nebula to the hydrogen ionizing
radiation \citep{Kj:89,Gv:00}. A similar result was also found 
for the LMC CSs in paper I.

\subsection{Luminosity-Nebular Radius and Surface Brightness Relations}
 
In Figure~2 we explore the relation between the CS luminosities derived in
this paper and the PN photometric radius (in {\rm pc}) from \cite{Setal:03}.
The CS and the nebula are interdependent systems, that is, the evolution of
the nebula is governed by the CS energetics, mainly through the stellar wind
and the ionizing radiation field, which are both a 
function of the CS mass and its evolutionary status.
Numerical simulations are required to understand how the evolution of the CS
luminosity relates to the nebular radius. Such models are
currently unavailable for the Magellanic Cloud PNe, however, as in Paper I,
we can obtain a qualitative indication of how the evolution of
nebular radius and CS luminosity relate to each other by using the models for
Galactic PNe of Villaver et al. (2002ab). In the Figure, the symbols
represents different morphologies 
(as in Fig.~1) and each line is labeled with the initial stellar mass used in
the simulation. 

In Figure~3 we explore the relation between the stellar luminosity and the
nebular surface brightness in the \hb~emission line (SB$_{H\beta}$, defined
as the integrated line flux divided by the nebular area $\pi$~R$_{phot}^2$;
where R$_{phot}$ is the photometric radius). As in Fig.~1 the 
symbols represent the different morphological types.
We find that low SB
objects are always located towards the position of the low luminosity CSs. The
relation is very similar to the one found for the LMC CSs in Paper I. The
3 objects located outside the main trend are (from the top left to the
right) SMP~25, MG~8 and MG~13.  This plot is a representation of the 
related fading of the CSs and the nebulae, as seen with {\it snapshots}
of the observed populations.

\subsection{Stellar Distribution on the Log L-Log T Plane, and the CSs
  Masses}    

In Figure~4 we show the distribution of the  detected CSs in the  Log L-Log T
plane. The evolutionary tracks for the SMC metallicity \citep{Vw:94}
encompass the observations well.  The CS masses listed in Table~4 have been
derived by comparing their location on the HR plane with the H-burning tracks
of \cite{Vw:94}, except in the case of J~18, whose mass was determined from
its position with respect to a He-burning track. Note that the CS
temperatures and luminosities (and therefore the masses) have been computed
within the assumption of a non-binary CS. The viability of the assumption
will be addressed in the following sub-section and in the discussion (\S5). We
have not derived masses for the two stars that lie below the evolutionary
tracks (SMP~11 and SMP~12).

In the cases where the helium flux at $\lambda$4686 was not available
(SMP~1, SMP~6, SMP~8, SMP~17, SMP~20, SMP~24, and SMP~27), we have used the
T$_Z$(H) to locate the CSs on the HR diagram. T$_Z$(H) underestimates the
temperature of the CSs hosted by optically thin PNe but it is accurate for
optically thick objects. Although the \heii~4686 fluxes of
SMP~1, SMP~6, SMP~17, and SMP~20 are only upper limits, and are 
a factor of 100 smaller than the \hb~flux, 
we still determine T$_Z$(\heii).
However, as the uncertainties are expected  to be very high, we have used
T$_Z$(H) to derive the masses. The difference in the CS masses 
derived using T$_Z$(\heii) or T$_Z$(H) is significant 
for SMP~1 and SMP~6, however, it is almost zero for SMP~24, and SMP~20.

\subsection{Constraints on a Potential Stellar Companion}
In the above, we implicitly assumed that all the light measured in the
stellar aperture is actually arising from the exciting
star. We do not have color 
information, and our magnitudes are measured through a broad-band filter. 
If the light in the aperture is contaminated by a
companion star (that is if the CS is embedded in a binary system) then   
the actual brightness of the PN progenitor in the passband would be 
fainter, which would mean that the true CS temperature and luminosity 
would be different from what was derived assuming no companion. In this
subsection we quantify the effects of hypothetical stellar companions,
subject to the strong constraint of a well determined distance, and examine
the consequences on the inferred temperature, luminosity, and mass of the PN 
progenitors.

A well determined distance modulus to the SMC allows strong 
constraints to be placed on the luminosity class of any binary 
companions to the CSs. Specifically, we can easily rule out 
supergiants (luminosity class I) as companions because such 
stars would be brighter in the STIS/50CCD bandpass than any 
CS in our sample. SdO Sub-dwarf companions are extremely unlikely 
because they evolve so quickly that they produce significant 
light in the V bandpasss (compared to the CSs) for only 
$\sim 10^5$ {\rm yr} after evolving of the AGB phase. Giant companions 
(luminosity class III) do produce the appropriate flux in our 
bandpass, so such companions cannot be ruled out. However, we 
believe it is statistically unlikely that many CSs in our small 
sample have Giant companions, given the brevity of this phase 
of evolution \citep{Rb:86}. \cite{Ytl:93} modeled a population 
of all binary stars and estimated the birthrates of binary CSs 
of different types. By this estimate, the birthrate of a binary 
CS formed by a CO white dwarf with a Giant companion is only 0.044 \% 
that of a CS formed by a CO white dwarf with a main-sequence 
companion. While there are a few PNs in the Galaxy with CS 
companions that are known to be Giants (e.g. A~35;\citealt{Jac:81}), 
we believe such cases are intrinsically rare and are unlikely 
to affect the conclusions of this paper. 

The combination of spectral type and luminosity class allows us to set 
strong constraints on the possibility of MS companions 
in our sample of CSs. We have recalculated the $V$ magnitudes of the 
CSs in Table~4 (those  with
derived masses) by assuming different MS binary companion contributions to
the light in the bandpass. By using {\bf synphot} we compute the instrumental
response to an input spectrum renormalized to the measured count rate. We
have considered a MS companion star that generates 50\%, 90\%, and 95\% of
the measured V-band flux (by using the synthetic spectral atlas from
Bruzual). Since the distance modulus for the SMC  is well known, the flux
limits severely limit the spectral type of the companion
\citep{Cox:00}. Basically, we adjust the $V$ magnitude of the model to match
the observed counts for our observing mode. Note that then the CS $V$
magnitude changes to get the measured count rate.

With this revised CS $V$ magnitude we recalculate the effective temperatures,
BCs, and CS luminosities following the procedure described in \S3.2. CS
masses have been derived as well based on the new position of the CSs in the
HR diagram (as in \S4.3). 

In Table~5 we give the recalculated CS parameters under different assumptions
about the relative  contribution  from a hypothetical MS companion
star. Column (1) lists the PN name, columns (2), (3), (4), and (5) we give
the assumed percentage of the companion to the count-rate, the $V$ magnitude
that the companion would have with the assumed contribution to the flux, the
spectral type that would correspond to this $V$ at the distance of the SMC
and the MS mass of the hypothetical companion from \cite{Cox:00}
respectively. Columns (6), (7), (8), (9), and (10) lists for the CS, $V$, the
effective temperature, the stellar luminosity (in Log), the CS mass and its
infered MS mass respectively. The  latter has  been derived from the initial-
to final-mass relation given in \cite{Vw:94} for CSs in the SMC. Obviously,
in those cases where the light is dominated by the true CS our mass
determinations in Table~4 will not be affected. Moreover, an arbitrarily
small contributions from the companion it is not worth considering because
then our estimate for the CS mass does not change meaningfully.

Comparison between columns (5) and (10) tell us whether a binary  MS
companion is physically reasonable. When the MS mass of the companion is
larger than the initial mass derived for the CS we fall into an evolutionary
contradiction: a less massive star cannot have evolved into a PN before its
more massive MS companion. Moreover, the difference in mass between the two
components of the system has to be large enough to allow one of them to
evolve into the PN domain while the other is still in the MS. This is correct
within the assumption that no significant mass-transfer has occured between
the CS progenitor and the postulated companion. PN formation in a close
binary system depends in detail upon the masses and structure of the
component stars at the time the (initially) more massive companion evolves
off of the MS \citep{Ibe:91}. A detailed interpretation in this context is
beyond the scope of this paper. From Table~5, by assuming that it will lead
to an evolutionary contradiction, we can safely rule out the possibility of
light contamination in all the objects, except in SMP~9, SMP~14, and
SMP~20. These three PNe deserve further investigation.

The ionization status of the nebula is a direct reflection of the effective
temperature of the CS. Zanstra analysis gives larger effective temperatures
when the CS  magnitude value increases for a given nebular flux. For SMP~9,
SMP~14, and SMP~20 we have checked for consistency between what would be the
CS effective temperatures admiting the possibility of a binary and the
nebular photoionizacion spectra.  We have computed the excitation class (EC)
of the nebula based on the definition  of \cite{Dm:90} and the nebular fluxes
in the literature (same references as in Table~2). We have used the relation
between EC and effective temperature of the CS given by \cite{Dm:91b}. This
relation is purely based on photoionization models and spectroscopic data for
the nebula. They are, then, safe from the photometric contamination that
might be affecting SMP~9, SMP~14, and SMP~20.

For SMP~14 and SMP~20 the measured light in the band-pass would have to be
completely dominated by the companion (above 90\%, see Table~5) which would
lead to effective temperatures above 200,000 \rm K. The ECs computed for the
nebula are 5 for SMP~14 , and 0 for SMP~20 corresponding to effective
temperatures of 100,000 and 40,000 \rm K respectively. Light contamination
from a binary  companion in our  photometry would imply CS temperatures are
inconsistent with the ionization status of these nebula. To stress this point
note also that the NeV~3426\AA~line have not been detected in either of these
PNs. This line being an additional indicator of EC above 8
(T$_{eff}\ge\sim$180,000 \rm K) \citep{Pot:84}.

A MS companion for SMP~9 emiting more than 65\% of the flux will move the CS
outside the PN tracks in the HR diagram. A CS temperature
consistent with the presence of a MS K0 or G8 companion (which will have to
contribute less than 65\% of the measured flux) will have to be above 230,000
\rm K to affect meaningfully the derived core mass. This temperature
correspond to an EC above 8 for the nebula which is inconsistent with
the EC derived from the nebular spectra, 5. The NeV~3426\AA~line have not
been detected in this nebula. We can not exclude, however, the presence of a
MS star contributing less than 50 \% of the measured light. Thus it is
possible that in SMP 9 the detected CS includes light from a companion star.
Note, however, that light contamination in this case would not meaningfully
affect the mass derived for the CS.

We conclude that we are likely not observing a companion star 
in any of the CSs in our sample, apart from the exceptions noted above, 
either because the flux from such companions would not be 
consistent with what is observed in our bandpass, or because 
the L, T, or the mass implied for a fainter CS would either 
be evolutionarily unreasonable or would be inconsistent with 
the excitation of the nebula, or because the likelihood of 
plausible companions is fairly small. It is difficult to apply 
the knowledge of the incidence of 
binary CSs in Galactic PNe, owing in part to the difficulty of 
demonstrating a large and volume-complete sample in the Galaxy 
\citep{Pot:96}. If we assume that the 10 \% 
fraction reported by \citep{Bon:00} for Galactic PNe also holds 
for the SMC, then we would expect two of the 21 positive 
detections in our sample would be in a close binary system. 
We have determined CS masses for only 14 objects, so it might 
be that one or two of the CSs actually has a close binary 
companion, which is consistent with what we found in the 
analysis above. Additional spectroscopy of the CSs 
in our sample would resolve the question. 

\subsection{Comparison of the SMC and LMC CSs masses}

The sample of SMC CSs analyzed in this paper and the sample of LMC CSs
observed with {\it STIS} in Paper I can be usefully compared for statistical
studies. The original samples were selected in a similar fashion, and were
observed with the same {\it HST} instrument and observing configuration.
The samples are of comparable size, and a similar fraction  of CSs were
detected within the observed PNe (67\% and 74\% for the SMC and LMC
respectively).

In Table~5 we give the basic statistics for the CSs masses, the effective
temperatures, and stellar luminosities of the SMC and LMC CSs.  
In column (1) we give the galaxy and, in parenthesis, the
sample size. Columns (2), and (3) show
the mean and the median of the mass distributions respectively. 
Columns (4), and (5) and (6), and (7) give similar statistics 
for the temperature and luminosities
distributions (in logarithmic scale). The numbers given in Table~5 also 
include SMP~1 and SMP~6, despite their masses having large errors 
(see the discussion in \S 5.3). 
If we exclude SMP~1 and SMP~6 from the SMC sample, the calculation of the 
mean mass only changes by 1$\%$ percent, and so does not affect our
conclusions.  

We find that the effective temperatures and luminosities (in logarithmic
scales) of the two CS samples are consistent with Gaussian distributions. The
CS masses, however, do not seem to be normally distributed. It is important
to note that our samples of CS masses have been determined free from 
distance uncertainties, thus their non-Gaussian 
distribution is noteworthy. 
In the theoretical PN luminosity function and other PN
applications, the CS mass distribution is generally assumed to be Gaussian.
We will examine the consequences of a non-Gaussian CS mass
distribution in a future paper.

In Figure~5 we show the normalized histogram of the mass distribution
obtained for the  SMC and the LMC samples (where the entire histograms have
been normalized to have a maximum value equal to 1). 
From Fig.~5 we can see
that the SMC hosts a larger fraction of low-mass CSs than the
LMC. The SMC sample lacks a population of stars between 0.65 and
0.75~\Mso~(see Table~4). We have tried to determine whether the mass
distributions of the 
LMC and SMC CSs are significantly different
by applying the Kolmogorov-Smirnov test (KS-test)
\citep{Ks:67}. The number of CSs masses is 14 and 16 for
the SMC and  LMC data respectively, and therefore the effective number
 of data points
(defined as the product of the number of data of each sample divided by its
sum) is 7.4, the KS-test is appropriate for an effective number of data
points greater than 4. 
The maximum difference, D, between the cumulative
distribution of the two datasets is 0.41 with a corresponding probability, P,
that the two distributions are the same of 0.12. P~=~1 indicates that two
distributions are identical. Figure 5 and the result of
the KS-test for the mass distributions, suggests that the mass
distribution of CSs in the LMC and SMC are different.

In Figure~6 we have plotted the SMC and the LMC CSs on the HR diagram, with
different symbols. It is apparent from  Fig.~6 (and from Table~5) that the
LMC CSs are hotter, and the stellar luminosities have a larger dispersion.
We have run a generalization of the KS test \citep{Ff:87, Pea:83} to test if
the 2-dimensional distribution in the logL-logT of the LMC and the SMC CSs
differs. We found D~=~0.34 and P~=~0.37, which is inconclusive in stating
that the locations in the HR diagram of the LMC and SMC CSs are drawn from
different distributions.

\section{SUMMARY AND DISCUSSION} 

We have obtained the physical parameters of 14 CSs in the SMC. We have
explored possible  correlations between the CS
(luminosity, effective temperature, and mass) and the nebula (photometric
radius and SB).  We find a similar pattern of evolution between the nebula and
the star as found for the LMC CSs in Paper I. We confirm that the stellar
luminosity evolves with nebular radius, as predicted by hydrodynamic
models \citep{Vmg:02}. We find a relation between the stellar luminosity and
the PN surface 
brightness in the SMC CSs, as was found in the LMC sample. We do not find any
significant relation between 
morphology and CS mass in the SMC sample, although it should be noted that
there are very few asymmetric PNe with detected CSs in the SMC, and generally
the undetected CSs will be the more massive. 

Given the additional restriction that a known distance imposes to the
measured flux we have explored the possibility that light from a
stellar companion is contaminating the photometric measurements. 
After a detailed analysis we were able to establish strong 
constraints on the existence and types of companions that 
would be consistent with the observed stellar and nebular 
fluxes. We showed that no CS could have a Supergiant 
companion, and we have argued that sdO sub-dwarf and Giant 
companions are very unlikely. We also showed that MS 
companions are are not consistent with the data and the 
implied evolutionary state of the PN progenitor, except for 
the case of SMP 9 where the feasible MS companion would not 
meaningfully affect the mass determined for the PN progenitor. 
It is important to note that our analysis does not exclude 
the possibility of a binary companion to the PN progenitor: 
we only limit or exclude the possibility of a companion as 
the source of the flux in the STIS/50CCD bandpass. 

We find that if the CS is evolving in a binary system the mass determined
within the assumption of a single CS will be an underestimation of the CS
mass if the CS is on a cooling track on the HR diagram. A lower (or the same)
mass will be determined if the CS is on the constant luminosity portion of
the track. A hotter CS results if we allow flux contamination from a binary
CS in the filter. While we cannot entirely rule out this possibility with
the data in hand, we have been allocated {\it HST} Cycle 13 orbits that will
allow us to increase the sample size and explore further its statistical
significance.

We find a difference in the mass distribution of the SMC and the LMC
CSs: the SMC has a narrower CS mass distribution than the LMC. Our findings
are  hampered by (1) the low number of SMC objects with
helium Zanstra temperature determinations; (2) the high  number of LMC CSs
masses determined 
from helium-burning tracks (resulting in slightly lower masses), and
(3) the overall small number of objects in the LMC and SMC samples. We should
note that the bias of point (2) above strengthens the conclusion that the SMC
has a narrower distribution of CS masses than the LMC. 

There are indications that the SMC PN population hosts slightly 
lower mass CSs compared to the LMC. 
If the initial mass distributions in the galaxies in
the 1-5 \Mso~range were the same, we would expect to find the opposite, that
is, higher final masses in the SMC compared to in the LMC, 
a consequence of the reduced mass-loss rate expected in a lower
metallicity environment. 
Since there is no evidence of a dependency of the IMF with metallicity
\citep{Sal:55, Sca:98}, we look to the history of star formation in the 
Magellanic Clouds to check whether we can expect differences in the final
mass distributions.

It seems well established that the star formation history of the LMC and SMC
are distinct \citep{Osm:96}. The LMC experienced an episode of star
formation $\sim$ 3-5 {\rm Gyr} ago \citep{Betal:92}, while the SMC seems to
have been forming stars at a constant rate during the last 2-12 {\rm Gyr}
\citep{Detal:01}. The stars with ages between 7.8$\times10^8$ and
2.7$\times10^8$ {\rm yr} are abundant in the LMC but are missing in the SMC. 
This latter population may  correspond to a burst of star formation in the
LMC. Thus, if the IMF is the same, then the star formation history
differences may account for the differences in the 
observed mass distribution.

\cite{Detal:85} found that the kinematics of the SMC PNe is that of a
spheroidal population without rotation. A study of Carbon stars, another
intermediate-age stellar population, affords a similar scenario 
\citep{Hsa:89, Hetal:97}, in agreement with the SMC PN population being
kinematically old. 

If we consider the initial-to-final mass relation and the evolutionary
timescales given by
\cite{Vw:94}, we find that our sample of CSs has an initial mass 
distribution that peaks at $\sim$ 1.5 \Mso, which would translate 
to an age of $\sim$3 {\rm Gyr}. 
SMP~25 and MG~8 are the most massive SMC CSs in our sample. Their
inferred initial masses (4.2 and 4.9 \Mso~respectively) 
sets a lower limit to
the age of the SMC PN population, $\sim 10^8$ {\rm yr}. These
two PNe are located in the same region of the SMC, the northeast region, 
where
\cite{Detal:85} found the kinematically younger PN population to be
concentrated. \cite{Cetal:01} found also that the eastern region of
the SMC, which faces the LMC, contains younger and more metal rich
clusters. Our estimated ages are then consistent with other studies of 
low- and intermediate-mass populations.

We cannot definitively claim that the difference found in the mass
distribution between the two Clouds is caused by differences in the
star formation histories, however we argue that it can explain it. A
metallicity dependency on the 
mass-loss rate alone, is not able to account for the observations. 

\acknowledgments  
We are very grateful to the referee, Stuart Pottash for his comments that
highly improve this paper. This work has been supported by NASA through grant
60-08271.01-97A from Space Telescope Science Institute, which is operated by
the Association of Universities for Research in Astronomy.

\begin{deluxetable}{lcc}
\tabletypesize{\scriptsize}
\tablenum{1}
\tablewidth{0pt}
\tablecaption{OBSERVATIONS}
\tablehead{
\multicolumn{1}{c}{}& 
\multicolumn{1}{c}{Integration} &
\multicolumn{1}{c}{CS}\\
\multicolumn{1}{c}{Name}& 
\multicolumn{1}{c}{time (s)} &
\multicolumn{1}{c}{Detection}\\
\multicolumn{1}{c}{(1)}& 
\multicolumn{1}{c}{(2)} & 
\multicolumn{1}{c}{(3)}}
\startdata
J~4           &  300    &  YES    \\
J~18          &  300    &  YES    \\
J~23          &  300    &  YES    \\
J~27          &  300    &  NO     \\
MA~1682       &  300    &  YES    \\
MA~1762       &  300    &  YES    \\
MG~8          &  120    &  YES    \\
MG~13         &  300    &  YES    \\
SMP~1         &  120    &  YES    \\
SMP~6         &  120    &  YES    \\
SMP~8         &  120    &  YES    \\
SMP~9         &  300    &  YES    \\
SMP~11        &  120    &  YES    \\
SMP~12        &  300    &  YES    \\
SMP~13        &  120    &  NO    \\
SMP~14        &  120    &  YES    \\
SMP~17        &  120    &  YES    \\
SMP~18        &  120    &  NO    \\
SMP~19        &  120    &  NO     \\
SMP~20        &  120    &  YES    \\
SMP~22        &  120    &  NO     \\
SMP~23        &  120    &  YES    \\
SMP~24        &  120    &  YES    \\ 
SMP~25        &  120    &  YES     \\ 
SMP~26        &  300    &  NO      \\
SMP~27        &  120    &  YES     \\ 
SP~34         &  300    &  NO      \\

\enddata
\end{deluxetable}


\begin{deluxetable}{lccccc}
\tabletypesize{\scriptsize}
\tablenum{2}
\tablewidth{0pt}
\tablecaption{\scriptsize MAGNITUDES, EXTINCTION, AND \heii~FLUXES}
\tablehead{
\multicolumn{1}{c}{Name}& 
\multicolumn{1}{c}{STMAG} &
\multicolumn{1}{c}{V} & 
\multicolumn{1}{c}{c} & 
\multicolumn{1}{c}{I(\heii)} & 
\multicolumn{1}{c}{Reference}\\
\multicolumn{1}{c}{(1)}& 
\multicolumn{1}{c}{(2)} &
\multicolumn{1}{c}{(3)} & 
\multicolumn{1}{c}{(4)} & 
\multicolumn{1}{c}{(5)} & 
\multicolumn{1}{c}{(6)}} 
\startdata
J~4\tablenotemark{a} &19.10$~\pm~$0.03&     18.81$~\pm~$0.07& 0.17 & 73.0$~\pm~$7.3 &  BL         \\
J~18 &20.93$~\pm~$0.04&     19.89$~\pm~$0.06& 0.42& 20.40$~\pm~$4.0 & JK93 \\  J~23 &20.17$~\pm~$0.04&     16.71$~\pm~$0.05& 1.65  & 57.0$~\pm~$5.7 &BL  \\
J~27 &$\ge$26.89&$\ge$25.89&0.41&  9.0$~\pm~$2.1 & BL  \\
MA~1682 &24.00$~\pm~$0.05&     24.30$~\pm~$0.10\tablenotemark{b}&\nodata& \nodata &  \nodata     \\ 
MA~1762 &23.86$~\pm~$0.08& 24.17$~\pm~$0.13\tablenotemark{b}&\nodata& \nodata        &  \nodata     \\
MG~8   &18.26$~\pm~$0.02& 18.09$~\pm~$0.06& 0.13 & 28.5$~\pm~$1.4 &LEI1\\
MG~13  &21.67$~\pm~$0.02& 21.03$~\pm~$0.05& 0.28  & 97.5$~\pm~$5.0 &  LEI1  \\
SMP~1  &17.65$~\pm~$0.02& 16.99$~\pm~$0.05& 0.29 &$\le$1.0 &  MEA\\
SMP~6  &17.66$~\pm~$0.12& 16.71$~\pm~$0.14& 0.39 &$\le$0.5 & LEI \\
SMP~8  &18.06$~\pm~$0.05& 18.27$~\pm~$0.10& 0.03 &  0.0   &  MON  \\
SMP~9  &24.37$~\pm~$0.41& 24.41$~\pm~$0.45& 0.07 & 59.6$~\pm~$3.0 &  LEI \\   
SMP~11 &20.25$~\pm~$0.14& 19.40$~\pm~$0.16& 0.35 &$\le$1.0&LEI1  \\
SMP~12 &19.70$~\pm~$0.04& 19.80$~\pm~$0.08& 0.06 &$\le$2.0  &LEI1 \\
SMP~13 &$\ge$17.17 & $\ge$16.87           & 0.19 &$\le$0.3&  LEI \\
SMP~14 &21.69$~\pm~$0.28& 21.74$~\pm~$0.32& 0.07 & 37.8$~\pm~$1.  &  LEI \\
SMP~17 &19.34$~\pm~$0.15& 19.41$~\pm~$0.19& 0.06 &$\le$1.0&  VAS \\
SMP~18\tablenotemark{a} &17.15$~\pm~$0.05&17.06$~\pm~$0.09 & 0.12 &0.0&  MON\\
SMP~19 &$\ge$20.79&$\ge$20.49 & 0.16 & 40.0$~\pm~$1.2 &  LEI \\
SMP~20 &20.30$~\pm~$0.02& 20.42$~\pm~$0.07& 0.05  &  $\le$0.3&  LEI\\
SMP~22 &$\ge$21.04 &$\ge$20.73& 0.17 & 60.4$~\pm~$1.2&LEI \\
SMP~23 &19.74$~\pm~$0.09& 19.68$~\pm~$0.13& 0.10 &  2.8$~\pm~$0.1 &  LEI \\   
SMP~24 &17.85$~\pm~$0.03& 17.98$~\pm~$0.07& 0.05 &  0.0           &  MON   \\
SMP~25 &18.65$~\pm~$0.01& 18.59$~\pm~$0.05& 0.10 & 53.3$~\pm~$3.2 &  LEI  \\
SMP~26 &$\ge$22.65 &$\ge$22.06 & 0.25 & 52.5$~\pm~$2.6&  LEI1  \\
SMP~27 &18.04$~\pm~$0.04& 18.20$~\pm~$0.09& 0.04 &  0.0       &  MON  \\
SP~34  &$\ge$23.08&$\ge$22.77& 0.16 & \nodata &\nodata \\

\enddata
\tablecomments{1-$\sigma$ errors are quoted throughout. 
The $\ge$ symbol refers to lower limit to the magnitude when
  the CS is not detected. } 
\tablenotetext{a}{Saturated data}
\tablenotetext{b}{No extinction constant available. The V mag 
was computed assuming  zero extinction} 
\tablerefs{(BL)\cite{BL}; (JK93)\cite{JK93};(LEI1) \cite{Ld:03}; (MEA)
  \cite{Dm:91a, Dm:91b}; 
  (LEI) \cite{Ld:96}; (VAS) \cite{VDM}; (MON) \cite{MBC}} 
\end{deluxetable}

\begin{deluxetable}{lccccccc}
\tabletypesize{\scriptsize}
\tablenum{3}
\tablewidth{0pt}
\tablecaption{\scriptsize CS PARAMETERS}
\tablehead{
\multicolumn{1}{c}{}& 
\multicolumn{1}{c}{T$_{\rm eff}$(He$_{\sc II}$)} & 
\multicolumn{1}{c}{T$_{\rm eff}$(H)}&
\multicolumn{1}{c}{}&
\multicolumn{1}{c}{}&
\multicolumn{1}{c}{}&
\multicolumn{1}{c}{}&
\multicolumn{1}{c}{}
\\
\multicolumn{1}{c}{NAME}& 
\multicolumn{1}{c}{(10$^3${\rm K})} & 
\multicolumn{1}{c}{(10$^3${\rm K})}&
\multicolumn{1}{c}{M$_{\rm V}$}&
\multicolumn{1}{c}{$\log L_*/L_{\sun}$}&
\multicolumn{1}{c}{BC}&
\multicolumn{1}{c}{M}&
\multicolumn{1}{c}{COMMENTS}\\
\multicolumn{1}{c}{(1)}& 
\multicolumn{1}{c}{(2)} & 
\multicolumn{1}{c}{(3)}&
\multicolumn{1}{c}{(4)}&
\multicolumn{1}{c}{(5)}&
\multicolumn{1}{c}{(6)}&
\multicolumn{1}{c}{(7)}&
\multicolumn{1}{c}{(8)}}
\startdata
J~4    &   74.1$\pm$5.1  & 29.0$\pm$3.2 & -0.02$\pm$0.07  & 4.17$\pm$0.09 &-5.65$\pm$0.20 & E& Saturated \\
J~18   &   52.6$\pm$2.3  & 20.2$\pm$1.3 &  1.06$\pm$0.06  & 3.33$\pm$0.06 &-4.63$\pm$0.13 &  R& \\
J~23   &   54.9$\pm$2.8  & 18.6$\pm$1.4 & -2.12$\pm$0.05  & 4.65$\pm$0.06 &-4.76$\pm$0.15 &U & Possibly not a PN\\
J~27   &   91.3$\pm$7.2  & 78.9$\pm$14.7& $\ge$ 7.03  &$\le$1.63          & -6.27$\pm$0.23  & B& Very faint object\\
MA~1682 &\nodata & \nodata & 5.47$\pm$0.1&\nodata & \nodata&B & no \hb fluxes, nor c available\\
MA~1762 &\nodata & \nodata & 5.34$\pm$0.13&\nodata & \nodata&E(bc) & no \hb fluxes, nor c available\\
MG~8   &   66.5$\pm$3.3  & 28.8$\pm$2.6 & -0.74$\pm$0.06  & 4.33$\pm$0.06 & -5.33$\pm$0.15  & E& \\
MG~13  &   98.3$\pm$8.7  & 43.7$\pm$6.8 &  2.20$\pm$0.05  & 3.62$\pm$0.11 & -6.49$\pm$0.26  & E& \\
SMP~1  &  $\le$ 46.4\tablenotemark{a}& 28.8$\pm$2.6 & -1.84$\pm$0.05  & 3.77$\pm$0.11 & -2.84$\pm$0.27  & U& \\
SMP~6  &  $\le$ 43.7\tablenotemark{a}& 28.2$\pm$2.5 & -2.11$\pm$0.14  & 3.86$\pm$0.12 & -2.78$\pm$0.26  & E& \\
SMP~8  &   \nodata       & 37.1$\pm$4.1 & -0.56$\pm$0.10  & 3.56$\pm$0.14 & -3.59$\pm$0.33 & R& \\
SMP~9  &  175.6$\pm$27.2 &156.8$\pm$44.8&  5.59$\pm$0.45  & 2.95$\pm$0.26
&-8.21$\pm$0.46 & R& Photometry in Nebula subs image\\
SMP~11 &  $\le$ 52.6\tablenotemark{a}& 40.9$\pm$5.0 &  0.57$\pm$0.16  & 3.22$\pm$0.16 & -3.88$\pm$0.36  &B&Photometry in Nebula subs image \\
SMP~12 &  $\le$ 51.7\tablenotemark{a}& 34.0$\pm$4.4 &  0.97$\pm$0.08  & 2.84$\pm$0.16 & -3.34$\pm$0.38 & E& \\
SMP~13 &  $\le$ 44.0\tablenotemark{a}& 31.3$\pm$2.2 & $\ge$-2.05  & $\le$4.03 & -3.09$\pm$0.21 & R& \\
SMP~14 &  116.8$\pm$11.4 & 83.5$\pm$17.6&  2.92$\pm$0.32  & 3.53$\pm$0.17 & -7.00$\pm$0.29  & R& \\
SMP~17 &  $\le$ 58.9\tablenotemark{a}& 58.4$\pm$7.3 &  0.58$\pm$0.19  & 3.65$\pm$0.17 & -4.94$\pm$0.37  &E&Photometry in Nebula subs image \\
SMP~18 &  \nodata     & 31.5$\pm$2.3 & -1.76$\pm$0.09 & 3.85$\pm$0.09 & -3.11$\pm$0.22 & U& Saturated\\
SMP~19 &  100.6$\pm$7.7  & 59.4$\pm$9.4 & $\ge$ 1.66 &$\le$3.90 & -6.56$\pm$0.23 & R& \\
SMP~20 &  $\le$ 58.5\tablenotemark{a}& 86.5$\pm$11.8&  1.59$\pm$0.07  & 3.71$\pm$0.16 & -6.11$\pm$0.40 & U&\\
SMP~22 &  122.2$\pm$14.5 & 76.7$\pm$18.2&$\ge$1.91  &$\le$ 4.04 & -7.14$\pm$0.35  & B& \\
SMP~23 &   62.1$\pm$3.0  & 41.6$\pm$5.1 &  0.85$\pm$0.13  & 3.61$\pm$0.08 & -5.12$\pm$0.14 & E(bc)& \\
SMP~24 &    \nodata      & 37.8$\pm$3.2 & -0.85$\pm$0.07  & 3.70$\pm$0.10 & -3.65$\pm$0.25 & E& \\
SMP~25 &   74.8$\pm$4.1  & 31.4$\pm$3.0 & -0.23$\pm$0.05  & 4.26$\pm$0.07 &-5.68$\pm$0.16 & E& \\
SMP~26 &  111.6$\pm$12.3 & 67.0$\pm$15.0&$\ge$  3.23  &$\le$ 3.41 & -6.87$\pm$0.33 & P& \\
SMP~27 &    \nodata      & 43.3$\pm$4.1 & -0.63$\pm$0.09  & 3.77$\pm$0.12 & -4.05$\pm$0.28  & R& \\
SP~34  &    \nodata      & 80.3$\pm$20.9&$\ge$ 3.94 &$\le$ 2.87 & -5.89$\pm$0.77 & R& \\
\enddata
\tablecomments{1-$\sigma$ errors are quoted throughout.
The $\ge$ symbol refers to the lower limit to the magnitude when
  the CS is not detected. In which case, the luminosities are upper
  limits and are preceded by a $\le$ symbol.} 
\tablenotetext{a}{Computed with the \heii~fluxes upper limits}
\end{deluxetable}

\begin{deluxetable}{lcc}
\tabletypesize{\scriptsize}
\tablenum{4}
\tablewidth{0pt}
\tablecaption{MASSES OF THE CENTRAL STARS}
\tablehead{
\multicolumn{1}{c}{Name}& 
\multicolumn{1}{c}{M [\Mso]} & 
\multicolumn{1}{c}{Comments}} 
\startdata
J~18        & 0.56    & He-burning track   \\
MG~8        & 0.88    & Core mass-Luminosity relation\\
MG~13       & 0.59    & Extrapolation H-burning track   \\
SMP~1       & 0.60\tablenotemark{a}    & Core mass-Luminosity relation   \\
SMP~6       & 0.63\tablenotemark{a}    & Core mass-Luminosity relation   \\
SMP~8       & 0.56\tablenotemark{a}    & Core mass-Luminosity relation   \\
SMP~9       & 0.67    & H-burning track   \\
SMP~14      & 0.59    & H-burning track  \\
SMP~17      & 0.59\tablenotemark{a}    & H-burning track  \\
SMP~20      & 0.59\tablenotemark{a}    & Core mass-Luminosity relation   \\
SMP~23      & 0.59    &  Extrapolation H-burning track  \\
SMP~24      & 0.59\tablenotemark{a}    & Core mass-Luminosity relation \\
SMP~25      & 0.82    &  Core mass-Luminosity relation   \\ 
SMP~27      & 0.60\tablenotemark{a}    & Core mass-Luminosity relation   \\
\enddata
\tablenotetext{a}{Derived from hydrogen Zanstra analysis and are 
therefore rather uncertain (see text). 
Note that the mass derived from the He-burning track
might be slightly lower than the mass derived from a H-burning track.} 
\end{deluxetable}

\begin{deluxetable}{lccccccccc}
\tabletypesize{\scriptsize}
\tablenum{5}
\tablewidth{0pt}
\tablecaption{BINARY COMPANION TO THE CS}
\tablehead{
\multicolumn{1}{c}{}& 
\multicolumn{4}{c}{Companion Star}&
\multicolumn{5}{c}{Central Star}\\
\multicolumn{1}{c}{Name}& 
\multicolumn{1}{c}{\% contribution}&
\multicolumn{1}{c}{$V$}&
\multicolumn{1}{c}{Sp type}&
\multicolumn{1}{c}{M [\Mso]}&
\multicolumn{1}{c}{$V$}&
\multicolumn{1}{c}{T$_{\rm eff}$(10$^3${\rm K})}&
\multicolumn{1}{c}{$\log L_*/L_{\sun}$}&
\multicolumn{1}{c}{M$_C$ [\Mso]}&
\multicolumn{1}{c}{M$_i$ [\Mso]}\\
\multicolumn{1}{c}{(1)}& 
\multicolumn{1}{c}{(2)}&
\multicolumn{1}{c}{(3)}&
\multicolumn{1}{c}{(4)}&
\multicolumn{1}{c}{(5)}&
\multicolumn{1}{c}{(6)}&
\multicolumn{1}{c}{(7)}&
\multicolumn{1}{c}{(8)}&
\multicolumn{1}{c}{(9)}&
\multicolumn{1}{c}{(10)}}
\startdata
J~18   & 50 &  20.10 &   A2 &  2.5 &  20.65 &  55.9  & 3.09  & \tablenotemark{a}&\nodata\\
       & 90 &  19.46 &   A1 &  2.7 &  22.44 &  65.5  & 2.57  & \tablenotemark{a}&\nodata\\
       & 95 &  19.40 &   A1 &  2.7 &  23.27 &  71.1  & 2.33  & \tablenotemark{a}&\nodata\\ 
MG~08  & 50 &  18.57 &   B8 &  3.0 &  19.02 &   73.1 &  4.06 &  0.69 &2.5\\
       & 90 &  17.93 &   B7 &  3.0 &  20.77 &   89.7 &  3.61 &  0.59 &1\\
       & 95 &  17.87 &   B7 &  3.0 &  21.52 &   99.2 &  3.42 &  0.57 &0.9\\
MG~13  & 50 &  21.17 &   A7 &  1.8 &  21.78 & 109.8 & 3.45 & 0.59 & 1\\
       & 90 &  20.53 &   A7 &  1.8 &  23.51 & 148.8 & 3.11 & 0.61 &1.2\\
       & 95 &  20.47 &   A7 &  1.8 &  24.36 & 178.7 & 2.99 & 0.67 &2\\
SMP~1  & 50 &  17.55 &   B3 &  7.6 &  17.75 &  33.2 & 3.64 & 0.59 &1\\
       & 90 &  16.92 &   B3 &  7.6 &  19.44 &  48.9 & 3.42 & 0.56 &0.89\\
       & 95 &  16.89 &   B3 &  7.6 &  19.70 &  52.4 & 3.40 & 0.56 &0.89\\
SMP~6  & 50 &  17.28 &   B3 &  7.6 &  17.46 &  32.4 & 3.72 & 0.59&1\\
       & 90 &  16.64 &   B3 &  7.6 &  19.21 &  48.0 & 3.49 & $\sim$0.57& $\sim$0.9\\
       & 95 &  16.58 &   B3 &  7.6 &  19.99 &  59.6 & 3.44 & $\sim$0.57& $\sim$0.9\\
SMP~8  & 50 &  18.68 &   B7 &  7.6 &  19.03 &  44.2 & 3.47 & 0.56&0.89\\
       & 90 &  18.05 &   B6 &  7.6 &  20.73 &  71.7 & 3.36 & $\sim$0.57&$\sim$0.9\\
       & 95 &  17.99 &   B6 &  7.6 &  21.46 &  92.4 & 3.37 & $\sim$0.57&$\sim$0.9\\
SMP~9  & 50 &  24.60 &   K0 & 0.79 &  25.16 & 210.9 & 2.87 & 0.70&2.5\\
       & 90 &  23.96 &   G8 & 0.85 &  26.91 & 363.9 & 2.82 & \tablenotemark{a}&\nodata\\
       & 95 &  23.90 &   G8 & 0.85 &  27.68 & 501.0 & 2.89 & \tablenotemark{a}&\nodata\\
SMP~14 & 50 &  21.89 &   F0 &  1.6 &  22.50 & 133.2 & 3.39 & 0.59&1\\
       & 90 &  21.25 &   F0 &  1.6 &  24.29 & 194.7 & 3.12 & 0.67&2\\
       & 95 &  21.20 &   F0 &  1.6 &  24.92 & 229.5 & 3.06 & 0.70&2.5\\
SMP~17 & 50 &  19.61 &   A0 &  2.9 &  20.17 &  74.1 & 3.62 & 0.59&1\\
       & 90 &  18.97 &   A0 &  2.9 &  21.95 & 144.9 & 3.71 & 0.64&1.5\\
       & 95 &  18.92 &   A0 &  2.9 &  22.59 & 190.1 & 3.77 & 0.68&2.2\\
SMP~20 & 50 &  20.57 &   A4 &  2.3 &  21.17 &  114.5& 3.74 & 0.64&1.5  \\
       & 90 &  19.93 &   A3 &  2.3 &  22.89 &  236.4& 3.91 & 0.69&2.5\\
       & 95 &  19.87 &   A3 &  2.3 &  23.64 &  331.0& 4.01 & 0.86&3.5\\
SMP~23 & 50 &  19.87 &   A0 &  2.9 &  20.44 &  66.8 & 3.39 &$\sim$0.57&$\sim$0.89\\
       & 90 &  19.24 &   A0 &  2.9 &  22.14 &  80.0 & 2.93 & 0.56&0.89 \\
       & 95 &  19.18 &   A0 &  2.9 &  22.88 &  87.4 & 2.74 & 0.56&0.89\\
SMP~24 & 50 &  18.39 &   B6 &  5.2 &  18.73 &  45.1 & 3.61 & 0.59&1\\
       & 90 &  17.75 &   B6 &  5.2 &  20.50 &  75.4 & 3.51 &$\sim$0.58&$\sim$1 \\
       & 95 &  17.69 &   B6 &  5.2 &  21.29 & 100.0 & 3.53 & 0.59&1\\
SMP~25 & 50 &  18.87 &   B8 &  3.8 &  19.35 &  81.5 & 4.06 & 0.69&2.5\\
       & 90 &  18.24 &   B8 &  3.8 &  21.06 & 101.8 & 3.64 & 0.61&1.2\\
       & 95 &  18.18 &   B8 &  3.8 &  21.79 & 113.7 & 3.48 & 0.59&1\\
SMP~27 & 50 &  18.58 &   B7 &  4.5 &  18.96 &  52.8 & 3.70 & 0.60&1.1 \\
       & 90 &  17.95 &   B7 &  4.5 &  20.66 &  90.9 & 3.67 & 0.60&1.1\\
       & 95 &  17.89 &   B7 &  4.5 &  21.39 & 119.9 & 3.71 & 0.63&1.5\\
\enddata
\tablenotetext{a}{Mass cannot computed, CS out of the evolutionary tracks.} 
\end{deluxetable}

\begin{deluxetable}{lcccccc}
\tabletypesize{\scriptsize}
\tablenum{5}
\tablewidth{0pt}
\tablecaption{STATISTICS}
\tablehead{
\multicolumn{1}{c}{}& 
\multicolumn{2}{c}{Mass [\Mso]}& 
\multicolumn{2}{c}{$\log T_{\rm eff}$ [K]} &
\multicolumn{2}{c}{$\log L_*/L_{\sun}$} \\
\multicolumn{1}{c}{galaxy}& 
\multicolumn{1}{l}{$<M>$}&
\multicolumn{1}{l}{Median}&
\multicolumn{1}{l}{$<\log T_{\rm eff}>$}&
\multicolumn{1}{l}{Median}&
\multicolumn{1}{l}{$<\log L_*/L_{\sun}>$}&
\multicolumn{1}{l}{Median}\\
\multicolumn{1}{c}{(1)}&
\multicolumn{1}{c}{(2)}&
\multicolumn{1}{c}{(3)}&
\multicolumn{1}{c}{(4)}&
\multicolumn{1}{l}{(5)}&
\multicolumn{1}{c}{(6)}&
\multicolumn{1}{c}{(7)}}
\startdata
SMC(14)  & 0.63 & 0.59 & 4.78$\pm$0.23 & 4.79 & 3.69$\pm$ 0.34 & 3.70  \\
LMC(16)  & 0.65 & 0.63 & 4.91$\pm$0.21 & 4.95 & 3.61$\pm$ 0.40 & 3.62  \\
\enddata
\tablecomments{1-$\sigma$ errors are given.}
\end{deluxetable}

\begin{figure}
\plotone{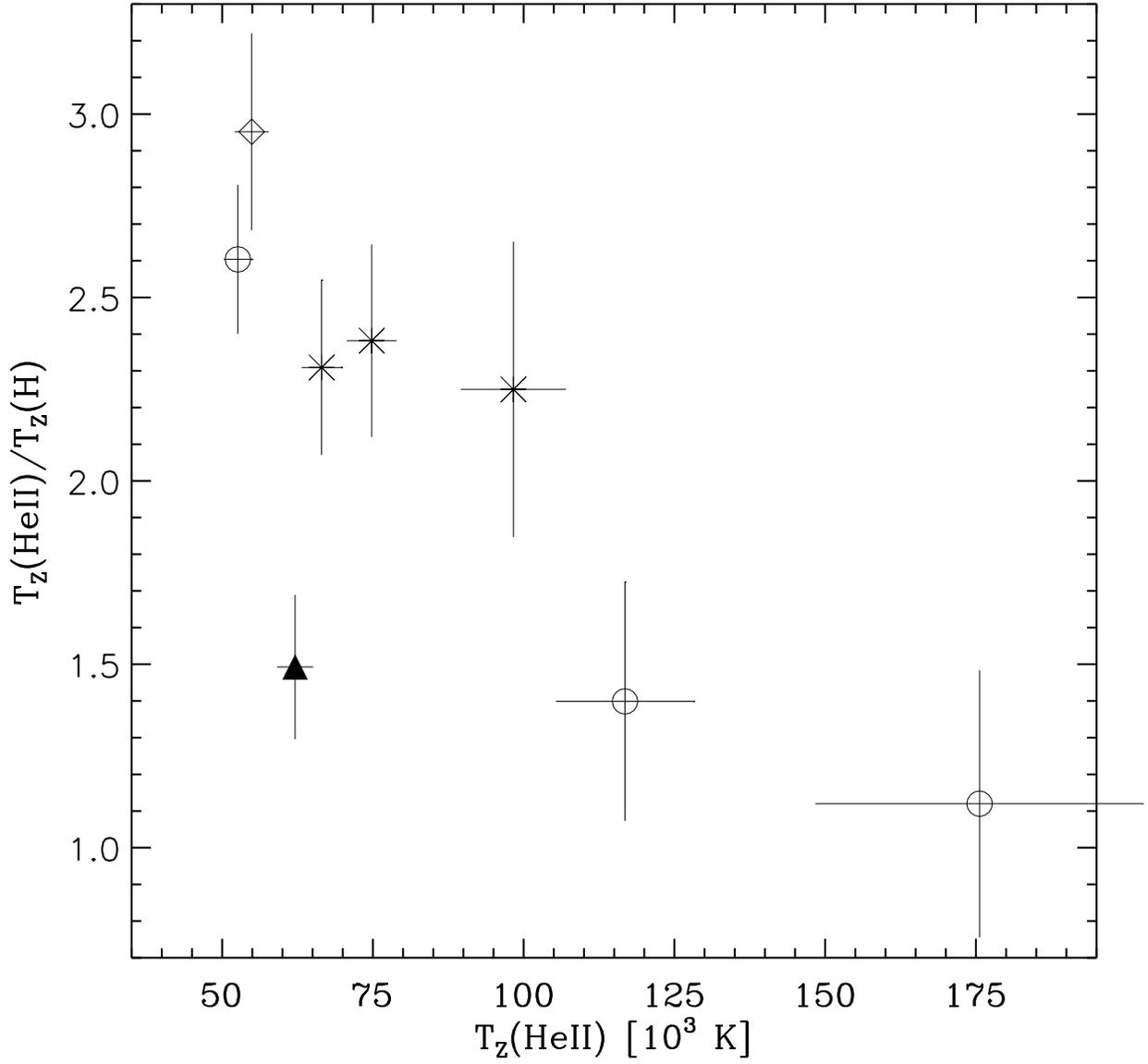}
\caption[ ]{Zanstra ratio versus He~{\sc ii} Zanstra
  temperature. The symbols represent the morphological types of the hosting
  nebulae: round (open
  circles), elliptical (asterisks), bipolar and quadrupolar (squares),
  bipolar core (triangles), point-symmetric (filled circles) and unresolved
  (diamonds).
\label{f1.eps}}
\end{figure}

\begin{figure}
\plotone{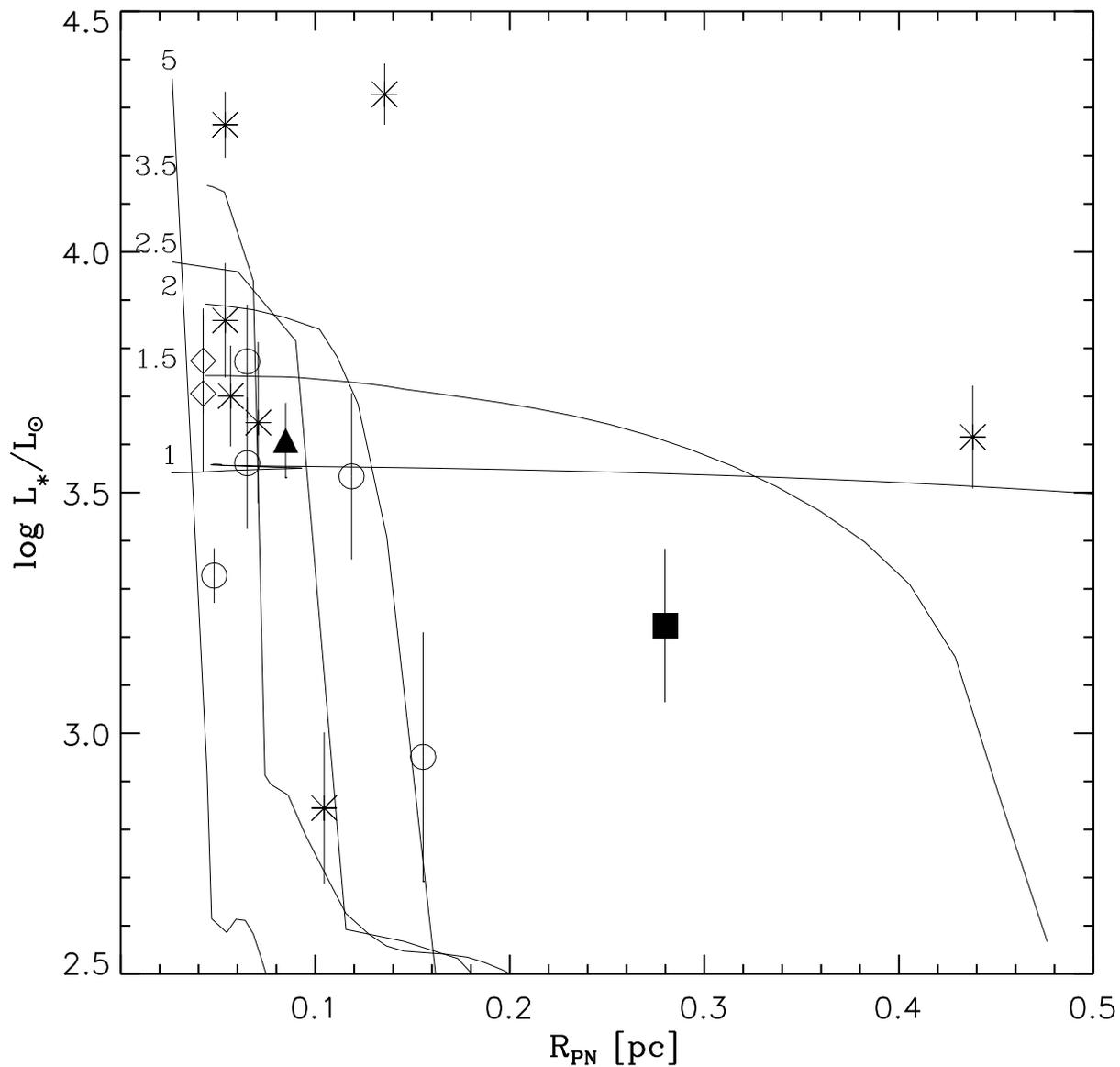}
\caption[ ]{The points represent the logarithm of the observed luminosity
  versus the physical radius 
  of the nebulae. The symbols are the same as used in Fig.~1. 
The solid line represent the
  evolution of the nebular radius versus the stellar luminosity taken from the
  numerical simulations of \cite{Vmg:02} for Galactic PN. 
Each line has been
  marked with the initial mass of the progenitor used in the hydrodynamical
  simulation.
\label{f2.eps}}
\end{figure}

\begin{figure}
\plotone{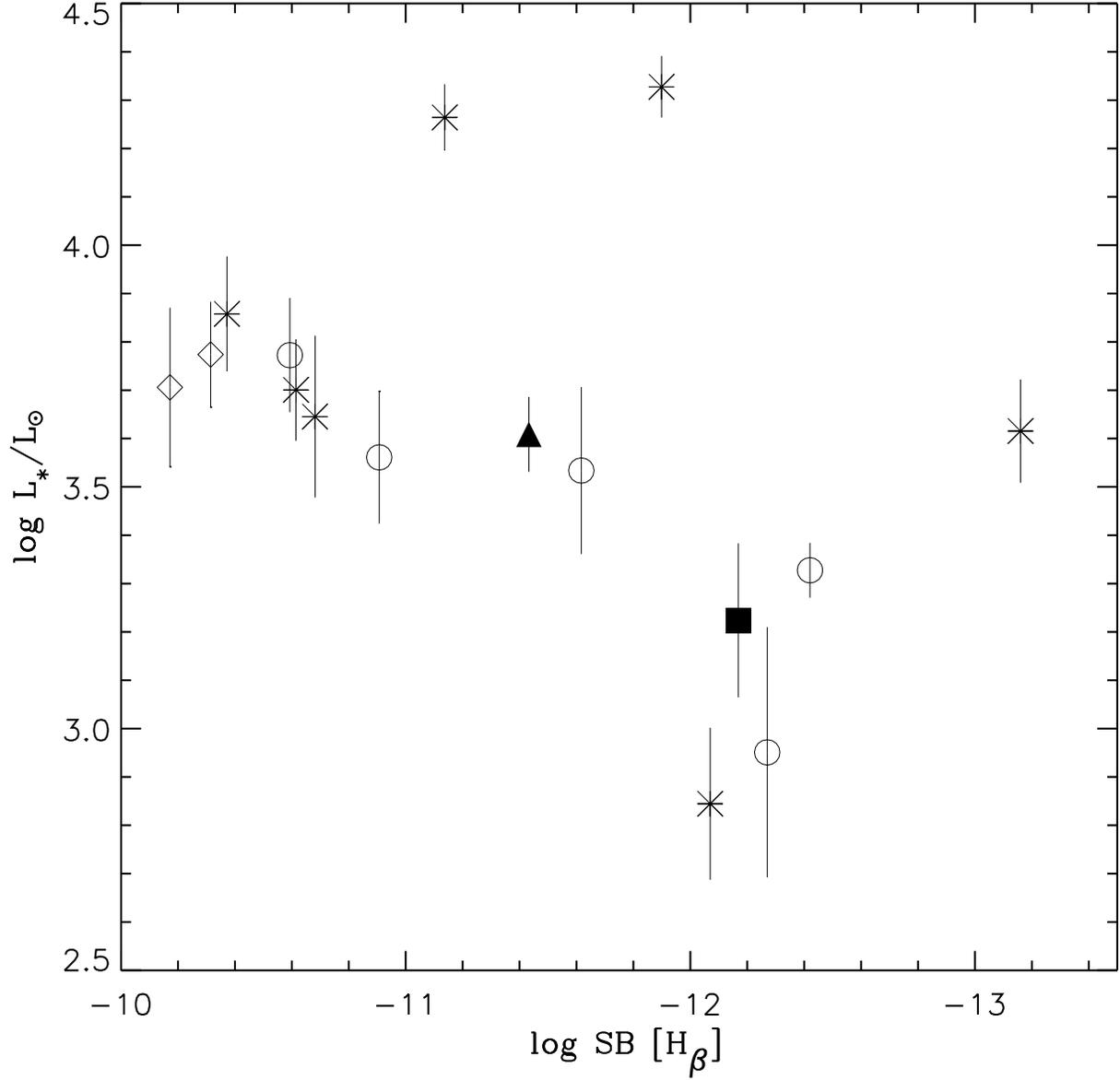}
\caption[ ]{Logarithm of the observed luminosity
  versus the surface brightness of the nebula in the \hb~lines. The
  symbols are the same as used in Fig.~1.
\label{f3.eps}}
\end{figure}

\begin{figure}
\plotone{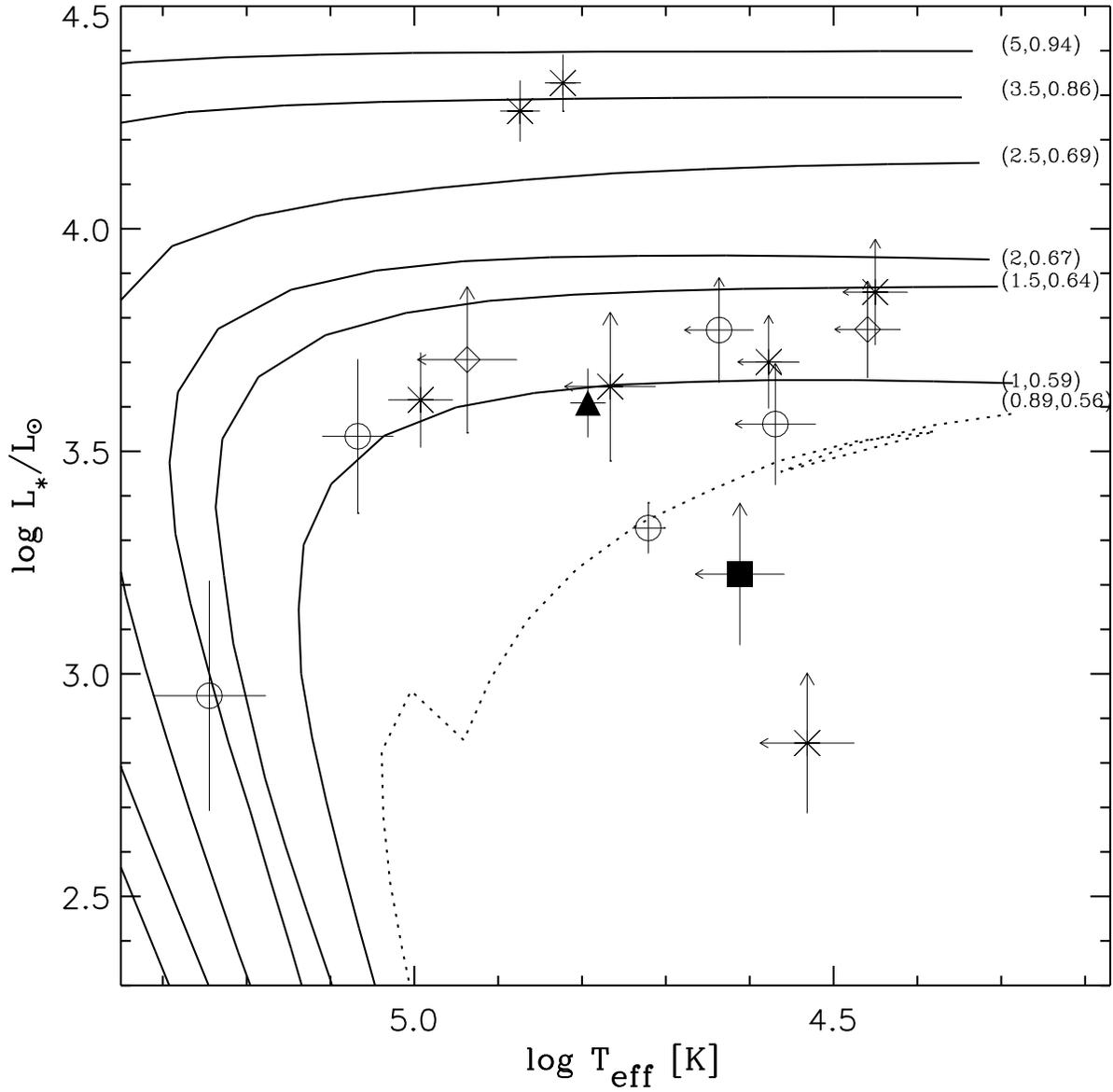}
\caption[ ]{HR diagram for the CSs. Symbols are the same as 
used in Fig.~1. We have maked with arrows 
those points for which we have used the H I Zanstra temperatures. 
Evolutionary tracks for SMC metallicities are from 
Vassiliadis \& Wood (1994). The initial and final masses are marked on
each track. The dotted lines is for a He-burnering track and the solid
lines for H-burners.
\label{f4.eps}}
\end{figure}

\begin{figure}
\plotone{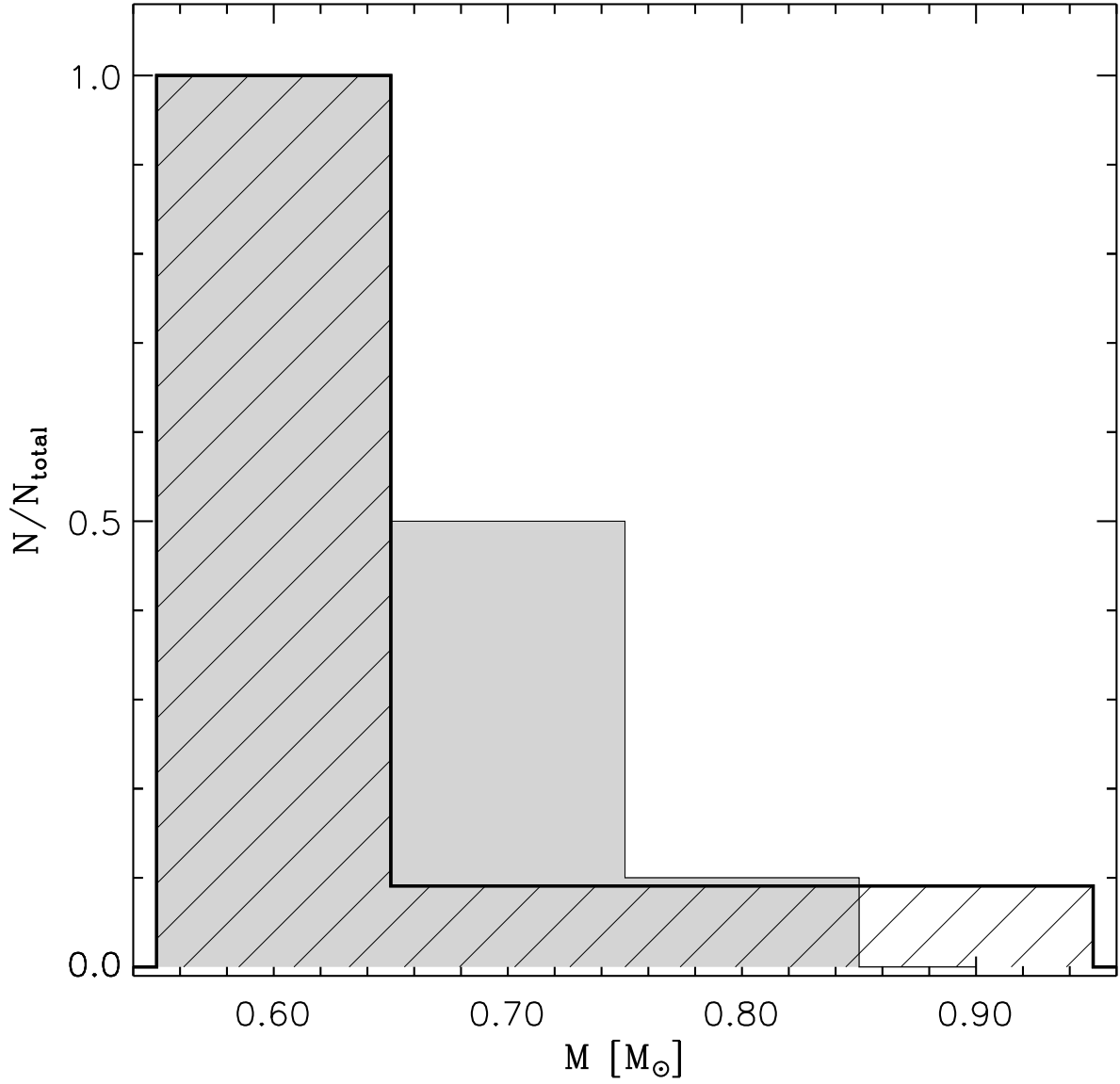}
\caption[ ]{Normalized mass distribution for the CSs in the SMC (dashed) and
  LMC (in gray).
\label{f5}}
\end{figure}

\begin{figure}
\plotone{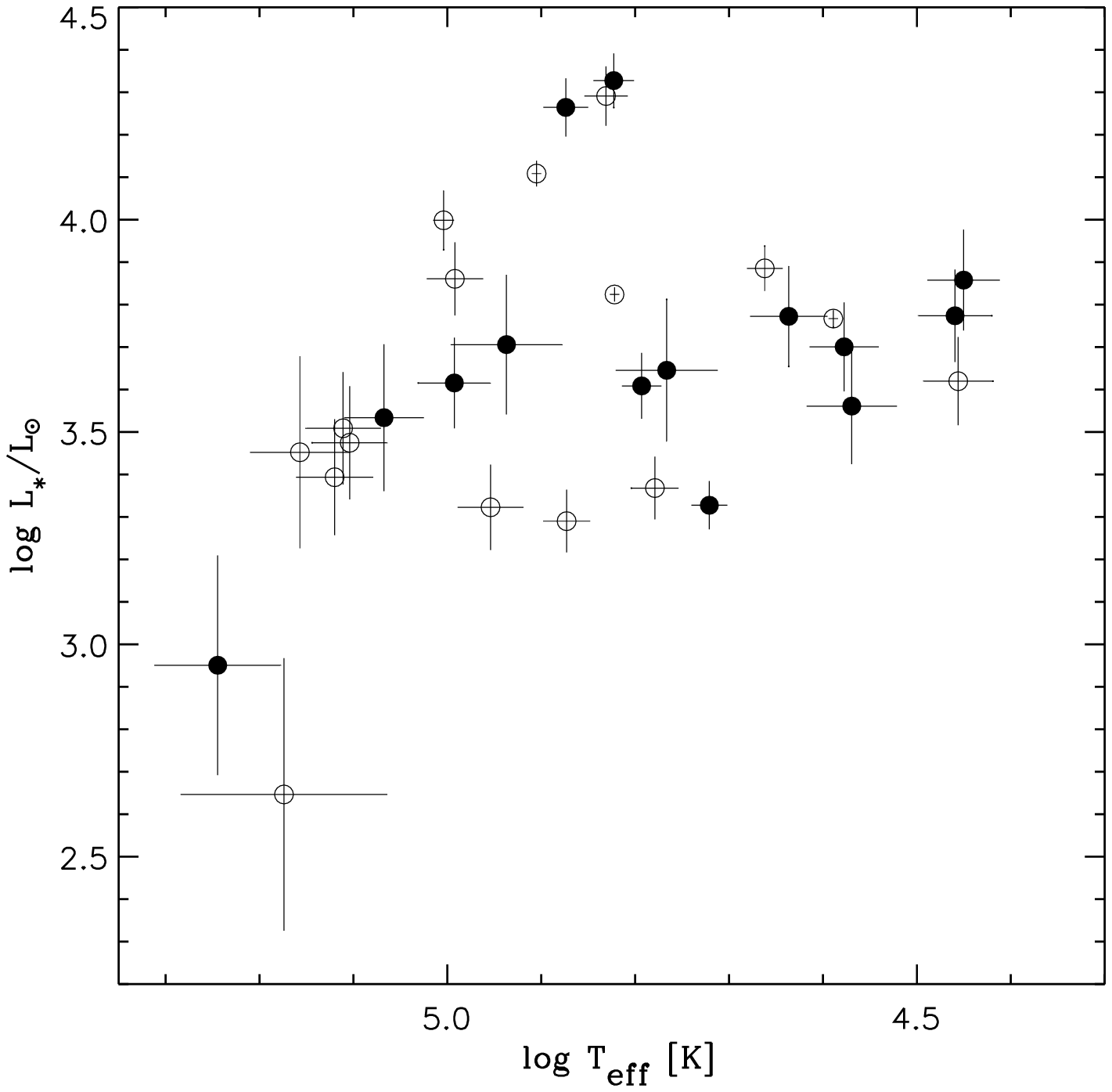}
\caption[ ]{HR diagram for the CSs in the SMC (filled circles) and LMC
  (open circles). Only those points for which the mass has been determined
  is plotted. 
\label{f6}}
\end{figure}


\begin{thebibliography}{}

\bibitem[Bond(2000)]{Bon:00} Bond, H.~E.\ 2000, ASP 
Conf.~Ser.~199: Asymmetrical Planetary Nebulae II: From Origins to 
Microstructures, 115 

\bibitem[Cox(2000)]{Cox:00} Cox, A.~N.\ 2000, Allen's 
astrophysical quantities, 4th ed.~ Publisher: New York: AIP Press; 
Springer, 2000.~ Editedy by Arthur N.~Cox.~ ISBN: 0387987460,  

\bibitem[Allen(1976)]{All:76} 
Allen, C.~W.\ 1976, Astrophysical Quantities, (London: Athlone)
\bibitem[Bertelli et al.(1992)]{Betal:92} 
Bertelli, G., Mateo, M., Chiosi, C., \& Bressan, A.\ 1992, \apj,
\bibitem[Boroson \& Liebert(1989)]{BL}
Boroson, T. A., \& Liebert, J. 1989, \apj, 339, 844 
\bibitem[Bowen(1988)]{Bow:88}
Bowen, G.~H.\ 1988, \apj, 329, 299 


\bibitem[Brown et al.(2002)]{Betal:02}
Brown, T. et al. 2002, ``{\it HST} STIS Data Handbook'', version 4.0, ed. 
B. Mobasher, (Baltimore:STScI) 

\bibitem[Chakravarti, Laha, \& Roy(1967)]{Ks:67}
Chakravarti, Laha, \& Roy, (1967). Handbook of Methods of
   Applied Statistics, Volume I, John Wiley and Sons, pp. 392-394.

\bibitem[Code et al.(1976)]{Cod:76} Code, 
A.~D., Bless, R.~C., Davis, J., \& Brown, R.~H.\ 1976, \apj, 203, 417 

\bibitem[Crowl et al.(2001)]{Cetal:01} Crowl, H.~H., Sarajedini, 
A., Piatti, A.~E., Geisler, D., Bica, E., Clari{\' a}, J.~J., \& Santos, 
J.~F.~C.\ 2001, \aj, 122, 220 

\bibitem[Dolphin et al.(2001)]{Detal:01} Dolphin, A.~E., Walker, 
A.~R., Hodge, P.~W., Mateo, M., Olszewski, E.~W., Schommer, R.~A., \& 
Suntzeff, N.~B.\ 2001, \apj, 562, 303 


\bibitem[Dominguez et al. (1999)]{Detal:99} 
Dominguez, I., Chieffi, A., Limongi, M., \& Straniero, O.\ 1999, \apj, 524,
226  

\bibitem[Dopita, Lawrence, Ford, \& Webster(1985)]{Detal:85} 
Dopita, M.~A., Lawrence, C.~J., Ford, H.~C., \& Webster, B.~L.\ 1985, \apj, 
296, 390 

\bibitem[Dopita \& Meatheringham(1990)]{Dm:90} Dopita, 
M.~A.~\& Meatheringham, S.~J.\ 1990, \apj, 357, 140 
\bibitem[Dopita \& Meatheringham(1991a)]{Dm:91a} Dopita, 
M.~A.~\& Meatheringham, S.~J.\ 1991a, \apj, 367, 115 

\bibitem[Dopita \& Meatheringham(1991b)]{Dm:91b} Dopita, 
M.~A.~\& Meatheringham, S.~J.\ 1991b, \apj, 377, 480 

\bibitem[Fasano \& Franceschini(1987)]{Ff:87}
Fasano, G. \& Franceschini, A.\ 1987, \mnras, 225, 155

\bibitem[Gardiner \& Hatzidimitriou(1992)]{Gh:92} Gardiner, 
L.~T.~\& Hatzidimitriou, D.\ 1992, \mnras, 257, 195 

\bibitem[Gardiner \& Hawkins(1991)]{Gh:91} Gardiner, L.~T.~\& Hawkins,
  M.~R.~S.\ 1991, \mnras, 251, 174 

\bibitem[Gilliland, Goudfrooij \& Kimble(1999)]{Ggk:99}
Gilliland, R. L., Goudfrooij, P. \& Kimble, R. A. 1999, PASP, 111, 1009

\bibitem[Girardi et al.(2000)]{Getal:00} 
Girardi, L., Bressan, A., Bertelli, G., \& Chiosi, C.\ 2000, \aaps, 141, 
371 

\bibitem[Gruenwald \& Viegas(2000)]{Gv:00} Gruenwald, R.~\& 
Viegas, S.~M.\ 2000, \apj, 543, 889 

\bibitem[Hardy, Suntzeff, \& Azzopardi(1989)]{Hsa:89} Hardy, 
E., Suntzeff, N.~B., \& Azzopardi, M.\ 1989, \apj, 344, 210 


\bibitem[Harman \& Seaton(1966)]{Hs:66} Harman, R.~F.~\& 
Seaton, M.~J.\ 1966, \mnras, 132, 15 

\bibitem[Hatzidimitriou \& Hawkins(1989)]{Hh:89} 
Hatzidimitriou, D.~\& Hawkins, M.~R.~S.\ 1989, \mnras, 241, 667 

\bibitem[Hatzidimitriou, Hawkins, \& 
Gyldenkerne(1989)]{Hhg:89} Hatzidimitriou, D., Hawkins, 
M.~R.~S., \& Gyldenkerne, K.\ 1989, \mnras, 241, 645 


\bibitem[Hatzidimitriou et al.(1997)]{Hetal:97} 
Hatzidimitriou, D., Croke, B.~F., Morgan, D.~H., \& Cannon, R.~D.\ 1997,
\aaps, 122, 507  

\bibitem[Howarth(1983)]{How:83} 
Howarth, I.~D.\ 1983, \mnras, 203, 301 

\bibitem[Iben(1991)]{Ibe:91} Iben, I.~J.\ 1991, \apjs, 76, 55 

\bibitem[Jacobi(1981)]{Jac:81}Jacoby, G. H. 1981, \apj, 244, 903

\bibitem[Jacoby \& Kaler(1993)]{JK93}
Jacoby, G. H., \& Kaler, J. B. 1993, \apj, 417, 209

\bibitem[Kaler(1983)]{Kal:83} 
Kaler, J.~B.\ 1983, \apj, 271, 188 

\bibitem[Kaler \& Jacoby(1989)]{Kj:89} Kaler, J.~B.~\& 
Jacoby, G.~H.\ 1989, \apj, 345, 871 

\bibitem[Leisy \& Dennefeld(1996)]{Ld:96} Leisy, P.~\& 
Dennefeld, M.\ 1996, \aaps, 116, 95 

\bibitem[Leisy \& Dennefeld(2003)]{Ld:03} Leisy, P.~\& 
Dennefeld, M.\ 2003, \aaps, in press

\bibitem[Leitherer et al.(2001)]{Letal:01}
Leitherer et al.(2001) ``STIS Instrument handbook'', version 5.1,
(Baltimore:STScI)

\bibitem[Monk, Barlow, \& Clegg(1988)]{MBC}
Monk, D. J., Barlow, M. J., \& Clegg, R. E. S. 1988, \mnras, 234, 583 

\bibitem[Olszewski, Suntzeff, \& Mateo(1996)]{Osm:96} 
Olszewski, E.~W., Suntzeff, N.~B., \& Mateo, M.\ 1996, \araa, 34, 511 

\bibitem[Osterbrock(1989)]{Ost:89}
Osterbrock, D. E. 1989, Astrophysics of Gaseous Nebulae and Active Galactic
Nuclei (Mill Valley: University Science Books) 

\bibitem[Peacock(1983)]{Pea:83}
Peacock, J.A. 1983, \mnras, 202, 615

\bibitem[Pottasch(1984)]{Pot:84}Pottasch, S. R. 1984, Planetary Nebulae
  (Dordrecht: Reidel) 
\bibitem[Pottasch(1996)]{Pot:96} Pottasch, S.~R.\ 1996, \aap, 
307, 561 

\bibitem[Rejkuba et al.(2000)]{Retal:00}
Rejkuba, M., Minniti, D., Gregg, M. D., Zijlstra, A. A., Alonso, M. V. \&
Goudfrooij, P., 2000, AJ, 120, 801 

\bibitem[Renzini \& Buzzoni(1986)]{Rb:86} Renzini, A.~\& 
Buzzoni, A.\ 1986, ASSL Vol.~122: Spectral Evolution of Galaxies, 195 

\bibitem[Russell \& Bessell(1989)]{Rb:89} Russell, S.~C.~\& 
Bessell, M.~S.\ 1989, \apjs, 70, 865 
\bibitem[Russell \& Dopita(1990)]{Rd:90} Russell, S.~C.~\& 
Dopita, M.~A.\ 1990, \apjs, 74, 93 

\bibitem[Salpeter(1955)]{Sal:55}
Salpeter, E.~E.\ 1955, \apj, 121, 161 


\bibitem[Seaton(1979)]{Sea:79} Seaton, M.~J.\ 1979, \mnras, 
187, 73P 

\bibitem[Savage \& Mathis(1979)]{Sm:79}
Savage, B. D. \& Mathis, J. S. 1979, ARA\&A 17, 73 

\bibitem[Scalo(1998)]{Sca:98} Scalo, J.\ 1998, ASP 
Conf.~Ser.~142: The Stellar Initial Mass Function (38th Herstmonceux 
Conference), 201
\bibitem[Stanghellini(2000)]{Sta:00} Stanghellini, L.\ 2000, 
\apss, 272, 181 

\bibitem[Stanghellini, et al.(2000)]{Setal:00}
Stanghellini, L., Shaw, R.~A., Balick, B., \& Blades, J.~C.~2000, \apjl, 534,
L167 

\bibitem[Stanghellini et al.(2002)]{Setal:02} 
Stanghellini, L., Shaw, R.~A., Mutchler, M., Palen, S., Balick, B., \&
Blades, J.~C.~2002, \apj, 575, 178 

\bibitem[Stanghellini et al.(2003)]{Setal:03} 
Stanghellini, L., Shaw, R.~A., Balick, B., Mutchler, M., Balick, B., \&
Blades, J.~C., Villaver, E.~2003, \apj, 596, 997

\bibitem[Stoy(1933)]{Sto:33} Stoy, R.~H.\ 1933, \mnras, 93, 
588 

\bibitem[Umeda et al.(1999)]{Uetal:99} 
Umeda, H., Nomoto, K., Yamaoka, H., \& Wanajo, S.\ 1999, \apj, 513, 861 

\bibitem[Vacca, Garmany, \& Shull(1996)]{Vgs:96} 
Vacca, W.~D., Garmany, C.~D., \& Shull, J.~M.\ 1996, \apj, 460, 914 

\bibitem[Vassiliadis et al.(1992)]{VDM}
Vassiliadis, E., Dopita, M. A., Morgan, D. H., \& Bell, J. F. 1992, \apjs,
83, 87  

\bibitem[Vassiliadis \& Wood(1994)]{Vw:94}
Vassiliadis, E.,\& Wood, P. R. 1994, \apjs, 92, 125

\bibitem[Vassiliadis et al.(1998)]{Vetal:98} Vassiliadis, E.~et 
al.\ 1998, \apj, 503, 253 
\bibitem[Villaver, Garc{\'{\i}}a-Segura, \& Manchado(2002a)]{Vgm:02} 
Villaver, E., Garc{\'{\i}}a-Segura, G., \& Manchado, A.\ 2002a, \apj, 571,
880  

\bibitem[Villaver, Manchado, \& Garc\'{\i}a-Segura(2002b)]{Vmg:02}
Villaver, E., Manchado, A., \& Garc\'{\i}a-Segura, G. 2002b, \apj, 581, 1204

\bibitem[Villaver, Stanghellini, \& Shaw(2003)]{Vss:03}
Villaver, E., Stanghellini, L., \& Shaw, R.~A.~2003, \apj, 597, 298 (Paper I)

\bibitem[Westerlund(1997)]{Wes:97}
Westerlund, B. E., 1997, The Magellanic Clouds. Cambridge
  Univ. Press, Cambridge

\bibitem[Willson, Bowen, \& Struck(1996)]{Wbs:96} Willson, 
L.~A., Bowen, G.~H., \& Struck, C.\ 1996, ASP Conf.~Ser.~ 98: From Stars to 
Galaxies: the Impact of Stellar Physics on Galaxy Evolution, 197 

\bibitem[Willson(2000)]{Wil:00} Willson, L.~A.\ 2000, \araa, 
38, 573 


\bibitem[Winters et al.(2000)]{Wetal:00} Winters, J.~M., Le 
Bertre, T., Jeong, K.~S., Helling, C., \& Sedlmayr, E.\ 2000, \aap, 361, 
641 
\bibitem[Wood(1979)]{Wod:79}
Wood, P. R. 1979, \apj, 227, 220

\bibitem[Yungelson, Tutukov, \& Livio(1993)]{Ytl:93} 
Yungelson, L.~R., Tutukov, A.~V., \& Livio, M.\ 1993, \apj, 418, 794 

\bibitem[Zanstra(1931)]{Zan:31} 
Zanstra, H. 1931, Publ. Dom. Astrophys. OBs. Victoria, 4, 209

\end{thebibliography}
\end{document}